\documentclass[letterpaper,twocolumn,10pt]{article}
\usepackage{usenix}
\usepackage{amsmath,amssymb,amsfonts,amsthm,mathtools}
\usepackage{graphicx}
\usepackage{booktabs}
\usepackage{microtype}
\usepackage{xcolor}
\usepackage{algorithm,algpseudocode}
\usepackage{url}
\usepackage{tabularx}

\hypersetup{hypertexnames=false}
\usepackage[skip=4pt]{caption}  

\newcommand{\sysname}{OctEx}

\graphicspath{{figures/}}

\title{\Large \bf \sysname: Billion-Scale Vector Search by Precomputed Lattice Neighbor Lists on GPUs}
\author{{\rm Dongfang Zhao (dzhao@uw.edu)}\\{\upshape\url{https://github.com/hpdic/octex}}}

\begin{document}

\maketitle

\begin{abstract}
Approximate nearest-neighbor search on GPUs serves retrieval-augmented generation, recommendation, and deduplication over datasets of up to billions of vectors. The fastest GPU indexes at high recall, such as CAGRA, walk proximity graphs in chained rounds, each of which waits for the distances of the previous one. High recall therefore costs many dependent rounds per query. Our insight is to spend the HBM of the GPUs on precomputed neighbor lists, so that a query reads its candidates in one parallel step instead of a chain of rounds. This paper turns the above insight into a GPU vector search system, namely \sysname, which holds hundreds of randomly projected copies of the dataset in GPU memory. Every copy rounds its vectors to the nearest points of a regular grid whose cells are nearly round and stores for every populated cell the list of vectors nearest to its center. A query reads the lists of its cells in all copies at once and then, in a fixed number of waves, the precomputed neighbor lists of its best hits. \sysname\ ranks the union of all hits by an int8 sketch before one exact rerank. \sysname\ runs the whole search of a batch as CUDA graphs over all GPUs of a server and places every part of its index in HBM or host memory by its access frequency. On four H100 GPUs at recall 0.95 to 0.995, \sysname\ outperforms CAGRA by 2.0$\times$ to 46.4$\times$ on seven datasets of up to one hundred million vectors, and by 1.46$\times$ to 2.05$\times$ on the one billion vectors of Deep1B and SIFT1B, whose indexes it builds in 44 to 47 minutes. \sysname\ also outperforms GGNN, PathWeaver, FlowANN, and the cuVS inverted file indexes IVF-PQ and IVF-RaBitQ by 5.4$\times$ to 343$\times$.
\end{abstract}

\section{Introduction}
\label{sec:intro}

Retrieval-augmented generation (RAG)~\cite{lewis2020rag}, dense passage retrieval such as DPR~\cite{karpukhin2020dpr} and ColBERT~\cite{khattab2020colbert}, and the deduplication of training data such as SemDeDup~\cite{abbas2023semdedup} require approximate nearest-neighbor (ANN) search at high recall, because every missed neighbor removes a relevant passage or keeps a duplicate. These applications search datasets of hundreds of millions to billions of embeddings, for instance the Deep1B benchmark~\cite{babenko2016deep} and the billion-scale challenge~\cite{simhadri2022bigann}. A modern GPU server holds several hundred gigabytes of HBM and tens of thousands of cores. The natural target for such a server is a search that turns this memory and this parallelism into throughput at high recall.

\begin{figure}[t]
\centering
\includegraphics[width=\columnwidth]{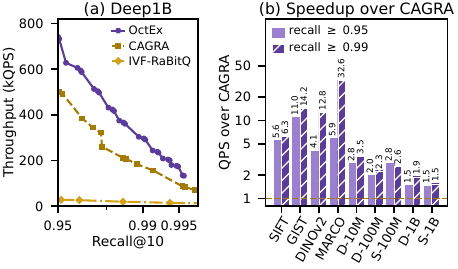}
\caption{\sysname\ against baselines on four H100 GPUs.}
\label{fig:opener}
\end{figure}

Existing GPU indexes follow two designs, of which proximity graphs pay for recall with a chain of dependent rounds and inverted file (IVF) indexes pay with scans of whole inverted lists. Proximity-graph indexes such as CAGRA~\cite{ootomo2024cagra}, GGNN~\cite{groh2023ggnn}, SONG~\cite{zhao2020song}, and BANG~\cite{venkatasubba2025bang} adapt CPU graphs such as HNSW~\cite{malkov2020hnsw}, NSG~\cite{fu2019nsg}, and DiskANN~\cite{subramanya2019diskann} to the GPU. Each of their iterations expands the best unvisited vertices, which depend on the distances of the previous iteration. A query thus advances through a chain of rounds whose length grows with the recall target. IVF indexes such as IVF-PQ in Faiss~\cite{johnson2019faiss} and cuVS~\cite{nvidia2025cuvs} and the recent IVF-RaBitQ~\cite{shi2026ivfrabitq} avoid the chain, as a query probes a fixed number of inverted lists in one pass. However, IVF indexes scan every compressed code of every probed list, which keeps their throughput on four H100 GPUs one to two orders of magnitude below CAGRA in our measurements. Figure~\ref{fig:opener}(a) quantifies the resulting cost on Deep1B, where CAGRA on four H100 GPUs falls from 156k queries per second (QPS) at recall 0.992 to 87k at recall 0.995 as each further 0.003 of recall needs a longer chain of rounds. On the same GPUs, IVF-RaBitQ remains below 28k QPS.

Breaking the chain requires candidates that do not depend on an earlier round. Precomputed lists of nearest vectors provide such candidates: a query that finds the list stored for its own position reads its neighbors in one step. Lists keyed by positions in the original space are, however, infeasible in hundreds of dimensions, because a ball around the query intersects almost every partition of the space before it reaches the tenth neighbor~\cite{beyer1999meaningful}. The lists must therefore be keyed by cells of low-dimensional projections and computed at build time, since an expansion at query time would reintroduce the chain.

Our insight is to precompute the nearest vectors per lattice cell and per random projection, so that a query reads them in parallel. This paper turns the above insight into a GPU vector search system, namely \sysname\ (Octa Explosion), which spends the HBM of the GPUs on these lists. \sysname\ projects the dataset into hundreds of random copies and rounds every block of eight projected coordinates to the nearest point of the $E_8$ lattice~\cite{conway1999splag}, a regular grid whose cells are the roundest known in eight dimensions. For every populated cell, \sysname\ stores the list of the vectors nearest to its center. At search time, a query decodes its own cell in every copy and reads the stored lists of these cells. The best hits become start points, whose precomputed neighbor lists the search reads in a fixed number of waves. Finally, \sysname\ ranks the union of all hits by an int8 sketch of every vector before it returns the top~$k$ by one exact rerank in half precision.

We implement \sysname\ for one server with four H100 GPUs, on which \sysname\ outperforms proximity-graph baselines (CAGRA~\cite{ootomo2024cagra}, GGNN~\cite{groh2023ggnn}, FlowANN~\cite{zhao2026flowann}, and PathWeaver~\cite{kim2025pathweaver}) and inverted file indexes (cuVS IVF-PQ~\cite{johnson2019faiss,nvidia2025cuvs} and IVF-RaBitQ~\cite{shi2026ivfrabitq}) on nine datasets from one million to one billion vectors. Figure~\ref{fig:opener}(b) presents the speedup over CAGRA across all nine datasets, which shows three results. First, the speedup of \sysname\ over CAGRA grows with the recall target, because \sysname\ keeps a fixed number of dependent steps while CAGRA needs a longer candidate list and more rounds per query. For example, on the MS MARCO passage embeddings that retrieval-augmented generation searches, the gain over CAGRA grows from 5.9$\times$ at recall 0.95 to 32.6$\times$ at recall 0.99. On ImageNet DINOv2, \sysname\ reaches recall 0.998, which CAGRA does not reach. Second, the gain holds on high-dimensional vectors, with 11.0$\times$ and 14.2$\times$ on GIST1M and 4.1$\times$ and 12.8$\times$ on DINOv2 at recall 0.95 and 0.99. Third, on the one billion vectors of Deep1B and SIFT1B, \sysname\ keeps the rows that every wave reads in HBM and moves only the half-precision vectors of the final rerank to host memory. There, \sysname\ outperforms CAGRA by 1.46$\times$ to 2.05$\times$ and cuVS IVF-RaBitQ by 20.2$\times$ to 38.9$\times$.

In summary, this work makes the following contributions:
\begin{itemize}
  \item We propose an ANN search by stored ripples, which replaces the chained rounds of graph walks with precomputed lists of nearest vectors at two kinds of sources. A query reads the ripples of its lattice fragments in hundreds of random projections in one parallel step. It then reads the neighbor rows of its best hits, with reverse edges, in a fixed number of waves (\S\ref{sec:design}).

  \item We implement \sysname\ as an open-source CUDA C++ module of cuVS~\cite{nvidia2025cuvs}, which builds the index of one billion vectors in 44 to 47 minutes by sorting instead of searching and runs the search of a batch as CUDA graphs over all GPUs of a server in one process. The release replaces the search calls of cuVS and Faiss in one line of Python (\S\ref{sec:impl}).
  
  \item We evaluate \sysname\ on nine datasets of up to one billion vectors against CAGRA, GGNN, FlowANN, PathWeaver, cuVS IVF-PQ, and IVF-RaBitQ on four H100 GPUs. \sysname\ outperforms CAGRA on every dataset at recall 0.95 to 0.995 and the other GPU graph systems GGNN, FlowANN, and PathWeaver by 5.4$\times$ to 27.9$\times$. Against the inverted file indexes cuVS IVF-PQ and IVF-RaBitQ, the gain reaches 7.0$\times$ to 343$\times$. At 100 to 1,000 queries per batch, \sysname\ answers a batch 2.2$\times$ to 6.0$\times$ faster than CAGRA on Deep100M and Deep1B. On one A100 of the previous generation, \sysname\ outperforms CAGRA on five datasets by 1.14$\times$ to 16.3$\times$ (\S\ref{sec:eval}).
\end{itemize}

\section{Background and Motivation}
\label{sec:motivation}

\subsection{GPU Vector Search}

The first GPU index family connects every vector to a few dozen neighbors in a proximity graph, in which a query starts from entry vertices and repeatedly expands the closest unvisited ones. CAGRA~\cite{ootomo2024cagra} builds a fixed-degree graph for the GPU and searches it with one thread block per query, which makes it the standard GPU baseline in cuVS~\cite{nvidia2025cuvs}. Other GPU graph indexes such as GGNN~\cite{groh2023ggnn}, SONG~\cite{zhao2020song}, and BANG~\cite{venkatasubba2025bang} follow the same expansion loop with different graphs and memory layouts. For billion-scale datasets, FlowANN~\cite{zhao2026flowann} extends CAGRA with a tiered graph that keeps part of the graph in host memory of one GPU server.

The second family, inverted file (IVF) indexes, clusters the dataset and scans the inverted lists of the clusters nearest to a query. The IVF-PQ index of Faiss~\cite{johnson2019faiss} and cuVS~\cite{nvidia2025cuvs} compresses every vector by product quantization~\cite{jegou2011pq}, whereas IVF-RaBitQ~\cite{shi2026ivfrabitq} replaces the codes by RaBitQ~\cite{gao2024rabitq} with an error bound per distance. Inverted multi-indexes~\cite{babenko2012imi} and optimized product quantization~\cite{ge2013opq} refine the coarse quantizer and the codes. In all of these indexes, a query reads every code of every probed cluster before an exact refinement of its candidates, which makes its cost grow with the cluster size rather than with the number of true neighbors.

\subsection{Chained Rounds of Graph Search}
\label{sec:chain}

The expansion loop of a graph search is a chain of dependent rounds, because the vertices that one round expands are the best results of the previous round. Although CAGRA computes the distances of all neighbors of the expanded vertices of one round in parallel, the next round can start only after a sort of the candidate list~\cite{ootomo2024cagra}. A higher recall target needs a longer candidate list and more rounds per query, while the GPU threads of a block wait at the end of every round for the slowest distance. Figure~\ref{fig:opener}(a) shows the consequence on Deep1B, where the throughput of CAGRA falls by 2.1$\times$ from recall 0.989 to 0.995.

The chain also limits the use of the memory of the server. A graph of one billion vertices with degree 32 occupies 119~GiB beside 179~GiB of half-precision vectors, which fills 298~GiB of the 372~GiB of HBM of four H100 GPUs with one copy of the CAGRA index. Additional memory cannot shorten the chain, as every round still depends on the last. In contrast, a search whose steps are independent could spend the memory on precomputed results and finish in a constant number of parallel steps per query.

\subsection{Search by Precomputed Lists}

A search with a constant number of steps needs candidates that do not depend on an earlier round. Precomputed lists of nearest vectors provide them, as long as a query finds its list by a computation on the query alone. A space partition gives such a computation, in which the query decodes its own cell and reads the list stored for that cell.

Because the volume of a ball in hundreds of dimensions concentrates near its surface~\cite{beyer1999meaningful}, the partition must be built in random low-dimensional projections. Random projections preserve distances in a few dozen dimensions up to a small distortion~\cite{johnson1984jl}. Within each projection, a lattice quantizer such as $E_8$ partitions the space into cells of nearly spherical shape~\cite{conway1999splag}. Formally, the $E_8$ lattice consists of points in $\mathbb{R}^8$ whose coordinates are either all integers or all half-integers and sum to an even integer: $E_8 = \{\mathbf{x} \in \mathbb{Z}^8 \cup (\mathbb{Z} + \frac{1}{2})^8 \mid \sum_{i=1}^8 x_i \in 2\mathbb{Z}\}$~\cite{conway1999splag}. Its Voronoi cells center the densest packing of equal balls in eight dimensions~\cite{conway1999splag,viazovska2017e8}. The vectors that share the cell of the query in a projection are therefore mostly near the query. Earlier CPU and disk indexes such as C2LSH~\cite{gan2012c2lsh}, QALSH~\cite{huang2015qalsh}, and SuCo~\cite{suco2025} enumerate the cell members of each random projection at query time and count them on the CPU. Their candidates are the members of the cells of a query, whose number varies from query to query and from cell to cell, whereas GPUs require fixed-shape parallel searches.

\subsection{Technical Challenges}
\label{sec:challenges}

Turning precomputed lists into a GPU search raises the following technical challenges:
\begin{itemize}
  \item Most cells that queries fall into are empty at fine spacing, as a product of multiple $E_8$ blocks has far more cells than a dataset has vectors. The lists must hence belong to cells that hold data and be found without a search for nearby cells at query time. Fragments at several spacings, each populated one with a stored fragment ripple, meet both conditions (\S\S\ref{sec:fragments} and~\ref{sec:ripples}).
  \item Part of the neighbors lie outside every cell of the query, a share that grows with the dataset, and must be reached without a chain of dependent rounds. Waves over neighbor rows, ranked by an int8 sketch, reach them in a fixed number of steps (\S\S\ref{sec:waves} and~\ref{sec:sketch}).
  \item The lists of all copies overlap and differ in length, so their union needs deduplication and ranking with few kernel launches per wave. Fixed-shape CUDA graphs and fused merge kernels provide both (\S\S\ref{sec:graph} and~\ref{sec:fused}).
  \item At one billion vectors, the projected copies, ranking sketches, neighbor lists, and raw vectors together exceed aggregate GPU HBM, while host memory serves random vector lookups at a small fraction of HBM bandwidth, and the lists of hundreds of copies threaten to make the build slow. \sysname\ places every part of its index in HBM or host memory by its access frequency (\S\S\ref{sec:multigpu} and~\ref{sec:tiers}) and builds all fragment ripples of a billion vectors in minutes by sorting the members of every fragment (\S\ref{sec:fastbuild}).
\end{itemize}

\section{Design}
\label{sec:design}

\subsection{System Overview}
\label{sec:overview}

\sysname\ indexes a dataset $X \subset \mathbb{R}^D$ through $C$ \emph{copies}. Copy $c$ maps every vector $\mathbf{x}$ by its own random projection, scaled by a \emph{spacing} $f$, to a point $\mathbf{y}$ of $\mathbb{R}^{8b}$, where $b$ is a small integer such as 4 (Section~\ref{sec:fragments} gives the map). The $8b$ coordinates of $\mathbf{y}$ split into $b$ blocks of eight. Each block rounds to its nearest point of the lattice $E_8$. We call the resulting tuple $(\mathbf{p}_1, \dots, \mathbf{p}_b) \in E_8^b$ the \emph{fragment} of $\mathbf{x}$ in copy $c$ at spacing $f$. It identifies the cell of $E_8^b$, a product of $b$ Voronoi cells of $E_8$, that contains $\mathbf{y}$. Two vectors share a fragment exactly when all $b$ of their blocks round to the same lattice points. A fragment is \emph{populated} when at least one vector of $X$ has it. Its \emph{center} is the point $(\mathbf{p}_1, \dots, \mathbf{p}_b)$ of $\mathbb{R}^{8b}$.

\sysname\ precomputes the near neighbors of two kinds of points as stored ripples. The \emph{stored ripple} of a source point is the list of the vectors of $X$ nearest to it, sorted by distance. A \emph{fragment ripple} has the center of a populated fragment as its source and measures distances in the projection of the copy. A query reads it through its own fragment in that copy. A \emph{vector ripple} has a vector of $X$ as its source and measures distances in $\mathbb{R}^D$. A query reads it through a vector that its search has already found. A search reads the fragment ripples of a query once, in its first round. Every later round is a \emph{wave}, in which each start point of the query reads its own vector ripple. The start points are the best candidates of the query so far. Both kinds of ripples are lists of vector identifiers that one GPU kernel gathers, which makes a search one read of fragment ripples followed by a few waves.

\begin{figure}[t]
\centering
\includegraphics[width=\columnwidth]{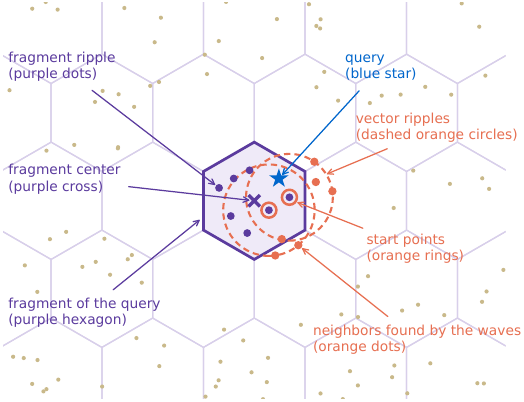}
\caption{The first round and the waves of one query.}
\label{fig:arch}
\end{figure}

A search of \sysname\ has four parameters $r$, $S$, $W$, and $R$, which the following steps define. Figure~\ref{fig:arch} follows one query through one of these copies. \sysname\ runs the same steps in all $C$ copies at once and for a whole batch of queries:
\begin{enumerate}
  \item \emph{Project.} One matrix multiplication maps the batch into the $C$ copies. In each copy, the query (blue star) falls into its fragment (purple hexagon) at several cell sizes, of which it uses the smallest one that has a stored list.
  \item \emph{Read the stored ripple.} The query reads the first $r$ entries of the fragment ripple of its fragment (purple dots), where the read depth $r$ is a search parameter. These entries are the $r$ vectors nearest to the fragment center (purple cross). The lists of all $C$ copies form the candidate pool of the query, from which a hash removes duplicates.
  \item \emph{Rank by sketch.} \sysname\ stores for every vector an int8 \emph{sketch} of its first principal coordinates and ranks the pool by sketch distance, without reading any full vector. The sketch distance of an entry is the distance between its sketch and the same principal coordinates of the query. The $S$ entries with the smallest sketch distances become the start points (orange rings), where $S$ is a search parameter.
  \item \emph{Waves.} In each wave, every start point reads its own vector ripple, which lists its nearest vectors in the full space (dashed orange circles). The waves thereby find neighbors outside the pool (orange dots). Vectors that the pool already holds keep their purple color inside the circles. The search ranks the new entries by sketch distance. The $S$ best entries whose rows no earlier wave has read become the start points of the next wave. The search runs a fixed number $W$ of waves.
  \item \emph{Rerank.} Sketch distances are approximate: the sketch keeps only the first principal coordinates of a vector in eight bits. After the last wave, \sysname\ therefore reads the full half-precision vectors of the $R$ entries with the smallest sketch distances, where the rerank depth $R$ is a search parameter. It then computes their exact distances to the query and returns the $k$ nearest of them. Since $R$ is a few dozen to a few hundred entries against thousands in the pool, the full vectors are read rarely enough to stay in host memory when they exceed HBM.
\end{enumerate}
Steps 1 to 3 form one parallel step, after which each wave adds one. A search thus waits for an earlier result $W$ times, where $W$ is a small constant, typically 1 to 6, that does not grow with the candidate pool. A graph walk instead waits once per round, for a number of rounds that grows with its candidate list and thus with the recall target (Section~\ref{sec:chain}).

The index that \sysname\ builds for a dataset $X$ consists of four parts: the keys and fragment ripples of the $C$ copies, the sketch rows, the vector ripples, and the half-precision vectors of $X$. A part that every query reads many times at random positions must stay in HBM, which serves random rows two orders of magnitude faster than host memory. A query reads $r$ entries of one fragment ripple in each of the $C$ copies, which makes $C r$ entries per query. The keys and fragment ripples of a copy therefore stay in the HBM of the GPU that owns the copy. A query also reads one sketch row for every distinct entry of its candidate pool. Each of the $W$ waves reads the vector ripples of $S$ start points, which makes $W S$ rows per query. For the same reason, the sketch rows and the vector ripples stay in HBM, split by rows over all GPUs when one GPU cannot hold them. We also call the vector ripples the neighbor rows. In contrast, the final rerank reads only $R$ half-precision vectors per query. When the dataset exceeds the HBM of the GPUs, the half-precision vectors consequently live in host memory.

\subsection{Lattice Fragments in Projected Copies}
\label{sec:fragments}

Each copy of the dataset projects every vector $\mathbf{x} \in \mathbb{R}^D$ into $8b$ dimensions and partitions the projection into products of $E_8$ cells. Specifically, copy $c$ draws a random orthonormal projection $U_c$ from the top principal subspace of the data and a uniform random shift $\mathbf{t}_c$ and maps $\mathbf{x}$ to $\mathbf{y} = (U_c^\top(\mathbf{x} - \boldsymbol{\mu}) + \mathbf{t}_c)/f$, where $\boldsymbol{\mu}$ denotes the data mean and $f$ the spacing in units of the projected standard deviation. The projection $\mathbf{y}$ splits into $b$ blocks of eight coordinates, each of which decodes to its nearest point of $E_8$ by the decoder of Conway and Sloane~\cite{conway1982e8decode}, which solves a special case of closest-point search in lattices~\cite{agrell2002closest}. The $b$ decoded points form the fragment of $\mathbf{x}$ in copy $c$ at spacing $f$, as Section~\ref{sec:overview} defines it.

The spacing controls the size of a fragment and with it the distance up to which vectors share the fragment of the query. A fine spacing gives small fragments that separate near from far vectors sharply but leave most queries in fragments without data. A coarse spacing gives fragments that almost always hold data but mix near and far vectors. To obtain both, \sysname\ decodes every vector at $L$ spacings $f_1 < \dots < f_L$, from which a query uses the finest spacing whose fragment holds a stored list.

The probability that a vector shares the fragment of the query decreases with their projected distance. Because the shift of a copy is uniform, the query falls at a uniform position within its own cell. A vector then shares the fragment of the query with the probability that this position stays inside the cell after a move by their projected difference, which is the share of the cell that overlaps its own translate by this difference. The cell of a lattice is convex, so this overlap can only shrink as the difference grows in any direction~\cite{conway1999splag}. The $b$ blocks of a fragment have independent shifts, which multiplies their probabilities.

A fragment therefore holds near vectors with higher probability than far vectors. Every copy draws its own rotation and shift, which makes the events of different copies independent. Over $C$ copies, the probability thus turns into an expected number of hits, which makes the union over copies contain the nearest vectors with a probability that rises with $C$.

\subsection{Stored Ripples}
\label{sec:ripples}

Because the fragment of a query holds no data in most copies at fine spacing, \sysname\ stores for every populated fragment the list of vectors nearest to its center, its fragment ripple. For each copy, spacing, and fragment that at least $p_{\min}$ vectors fall into, the build computes the list of the $l$ vectors nearest to the center of the fragment in the projection of the copy, sorted by distance. A query that falls into a stored fragment reads the first $r \leq l$ entries of its list. A query whose fragment is not stored at the finest spacing instead moves to the next coarser spacing, where fragments hold more data.

The stored list approximates the list of the vectors nearest to the query, given that the query lies near the center of its fragment. The approximation weakens as the offset $\epsilon$ of the query from the center grows. Let $\tau_r$ denote the projected distance of the $r$-th entry of the list to the center. By the triangle inequality, a vector whose projected distance to the query is below $\tau_r - \epsilon$ lies closer to the center than $\tau_r$, so that the first $r$ entries contain every such vector and cover this radius in full.

The reach $\tau_r - \epsilon$ of a list depends on the spacing and the list length, while the pruning threshold $p_{\min}$ sets its memory. The offset $\epsilon$ is at most the covering radius of the fragment, which is the largest distance from a point of its cell to its center. For a product of $b$ cells of $E_8$, this radius is $\sqrt{b}$, since the covering radius of $E_8$ is 1~\cite{conway1999splag}. Before the division by the spacing $f$, the offset in the projection is thus at most $f\sqrt{b}$, which a finer spacing reduces. A longer list instead increases $\tau_r$. The pruning threshold costs little coverage, since a fragment that holds few vectors also receives few queries when the queries follow the data distribution. As a result, dropping such a fragment below $p_{\min}$ moves only those queries to the next spacing, at a small loss of recall.

\subsection{Waves over Neighbor Rows}
\label{sec:waves}

\sysname\ reaches the neighbors that the first round misses by waves, each of which adds one dependent step. The first round of stored ripples finds most of the nearest neighbors but misses those that fall into other fragments in almost every copy. The vector ripples of Section~\ref{sec:overview} serve this purpose: the neighbor row of every vector lists its nearest vectors in the full space. In each wave, the $S$ best entries found so far read their neighbor rows. Since a wave reads $S$ rows in parallel and depends only on the ranking of the previous wave, a search with $W$ waves has $W + 1$ dependent steps in total, however long its candidate pool.

A neighbor row that lists only the nearest neighbors of its owner misses vectors whose own neighbors are near the owner but which no near vector lists. \sysname\ therefore builds each row from $k_f$ forward neighbors, up to $k_r$ reverse entries, and the remaining forward neighbors, where a reverse entry of $\mathbf{u}$ is a vector $\mathbf{v}$ that lists $\mathbf{u}$ among its first $k_s$ forward neighbors. Reverse entries give every vector a path back from the vectors that consider it near, along which a wave reaches every vector from the start points near it.

From one hundred million vectors on, \sysname\ builds the neighbor rows by its own search, while smaller datasets use exact rows. The first version of the rows comes from the first round alone, with every base vector as a query and the vector itself removed. Each later version runs the full search with waves over the reverse rows of the previous version, which raises the quality of the rows with each refinement.

\subsection{Ranking without Full Vectors}
\label{sec:sketch}

\sysname\ ranks the union of a query by an int8 sketch: exact distances for its thousands of distinct entries would read thousands of full vectors per query. The sketch stores for every vector its first principal coordinates in eight bits with one scale per coordinate. Each sketch distance reads one row of 64 to 128 bytes. The exact rerank thus needs only a few dozen entries whenever the sketch ranks the true neighbors among its best entries. The best $S$ entries by sketch distance become the start points. Only the best $R$ entries after the last wave receive exact half-precision distances, or in every wave when the vectors fit in HBM.

The separation of ranking from exact distances determines where the index can live. The sketch rows are read for every distinct entry and must stay in HBM. In contrast, the full vectors are read $R$ times per query and can live in host memory. Algorithm~\ref{alg:search} summarizes the search of one query with both kinds of ripples and both rankings.

\begin{algorithm}[t]
\caption{Search of one query $\mathbf{q}$ in \sysname.}
\label{alg:search}
\begin{algorithmic}[1]
\Require copies $1..C$, spacings $f_1 < \dots < f_L$, read depth $r$
\Require start points $S$, waves $W$, rerank depth $R$
\Ensure the approximate $k$ nearest neighbors of $\mathbf{q}$
\State $\mathcal{U} \gets \emptyset$
\For{$c = 1..C$ \textbf{in parallel}}
  \State $j \gets$ the finest spacing whose fragment of $\mathbf{q}$ is stored
  \State $\mathcal{U} \gets \mathcal{U} \cup$ the first $r$ entries of the list of that fragment
\EndFor
\State $\mathcal{T} \gets$ the $S$ entries of $\mathcal{U}$ nearest to $\mathbf{q}$ by sketch distance
\For{$w = 1..W$}
  \State $\mathcal{U} \gets \bigcup_{\mathbf{s} \in \mathcal{T}}$ the neighbor row of $\mathbf{s}$, in parallel
  \State $\mathcal{T} \gets$ the $S$ unread entries of $\mathcal{T} \cup \mathcal{U}$ nearest by sketch distance
\EndFor
\State \Return the exact top $k$ among the best $R$ of $\mathcal{T}$
\end{algorithmic}
\end{algorithm}

\paragraph{Time and space complexity.} A query of Algorithm~\ref{alg:search} performs $O(C r + W S L_t)$ memory reads and sketch distances and $O(R)$ exact distances in $W + 1$ dependent steps, where $L_t$ denotes the length of a neighbor row. Specifically, the first round reads $C r$ list entries and computes at most $C r$ sketch distances per query; similarly, each wave reads $S$ rows of length $L_t$ and computes at most $S L_t$ sketch distances. The index stores, per copy, $l$ identifiers for each of its stored fragments together with their keys; in addition, the shared part stores one sketch row and one neighbor row of $L_t$ identifiers per vector. The deduplication hash set of a query holds at most $\max(C r, S L_t)$ entries in any round.

\section{Implementation}
\label{sec:impl}

The implementation of \sysname\ consists of 11,425 lines of code. At its core, a CUDA C++ library of 5,452 lines forms the module \texttt{cuvs::neighbors::octex} of cuVS~\cite{nvidia2025cuvs}. Its search path takes 2,716 lines with 10 CUDA kernels and covers the loading of an index onto the GPUs and the search of a batch of queries. The builder takes another 2,736 lines with 16 kernels and computes the copies, the fragment ripples, the sketch, and the neighbor rows on the GPUs, with the radix sort of CUB for the fragment ripples of billion-vector datasets and the merges, and the IVF-Flat index of cuVS for the ripples of smaller ones. The C and Python interfaces take 701 and 1,322 lines. Unit tests take 1,421 lines. Command-line tools, examples, and further tests take the remaining 2,529 lines.

The source code of \sysname\ is available on GitHub\footnote{\url{https://github.com/hpdic/octex}}. The Python package is published on PyPI as \texttt{octex}\footnote{\url{https://pypi.org/project/octex}}. The release offers five interfaces:
{\raggedright
\begin{itemize}
  \item \emph{C++:} \texttt{build}, \texttt{load}, and \texttt{search} in the style of cuVS, with queries in host or device memory.
  \item \emph{C:} \texttt{cuvsOctexBuild}, \texttt{cuvsOctexLoad}, and \texttt{cuvsOctexSearch} on DLPack tensors, following the C API of cuVS.
  \item \emph{Python:} \texttt{octex.build}, \texttt{octex.load}, and \texttt{Index.search} on NumPy, CuPy, or PyTorch arrays, or any array with DLPack or the CUDA array interface, without copies.
  \item \emph{Drop-ins:} \texttt{from octex import cuvs as cagra} for the CAGRA calls of cuVS and \texttt{from octex import faiss} for the GPU index calls of Faiss, each replacing one import in existing code.
  \item \emph{Command line:} \texttt{octex\_build} and \texttt{octex\_bench}, which build indexes and run the measurements of Section~\ref{sec:eval}.
\end{itemize}
\par}

\subsection{One CUDA Graph per Batch}
\label{sec:graph}

\sysname\ captures the search of a batch on each GPU as CUDA graphs~\cite{nvidia2025cudagraphs}, one for the first round and one for the exchange of start points and all waves. Every kernel of the graph has a fixed shape: the projection is one matrix multiplication over all copies; similarly, the decoding, gather, and rerank kernels assign a fixed number of threads to every query. Queries that need fewer entries than the maximum skip work inside the kernels instead of changing the launch shape, which spares the host any synchronization inside a search. Every step is a CUDA kernel or a cuBLAS call in C++, with no framework in the search path.

Every per-query structure inside the graph takes a size that follows from the maximum number of entries of the query. The deduplication of a union uses an open-addressing hash set in the shared memory of the block~\cite{ashkiani2018hashtable}. Such a set is one fixed array, in which an entry that finds its slot taken moves on to the next free slot. Its size is the next power of two above 1.5 times the maximum number of entries of the query, at most 32,768 slots. The gather kernel inserts the list entries of a query into this set and splits them evenly over the threads of the block by a prefix sum of the list lengths. One lane per 16 sketch coordinates of each new entry then computes its sketch distance. Similarly, the kernel that keeps the $S$ start points sorts the next power of two above $S + R$ keys of distance and identifier by a bitonic network~\cite{batcher1968sorting}, instead of a fixed size of 1,024 keys. The exact rerank selects the $R$ best entries by a radix selection over their sketch distances and reads their half-precision rows.

The graphs keep every GPU busy through a batch. In a trace of one batch of 10,000 queries on Deep1B over four H100 GPUs, the kernels of the graphs leave each GPU idle for at most 0.06~ms in total, against 0.17 to 0.25~ms when the host launches every kernel. The four GPUs finish the batch within 0.2~ms of each other.

\subsection{Fused Merge Kernels}
\label{sec:fused}

Two fused kernels merge the start points of a query and select those of the next wave, each in one thread block per query. The merge kernel radix-selects the $S$ sketch-nearest of the new entries of a query in one thread block and then sorts their keys of distance and identifier together with the kept start points by a bitonic network in shared memory. Because the same identifier always carries the same sketch distance, a repeated identifier becomes an equal key and collapses during the sort. The selection kernel inserts the identifiers that earlier waves expanded into a hash set in shared memory and takes the first $S$ kept entries, in order, that the set does not contain.

Both kernels return the same start points as a merge by sorting, up to the order among equal sketch distances. Since each kernel keeps its intermediate keys in shared memory, a wave issues one launch for the merge and one for the selection per batch. The merge also removes the duplicates that the start points of successive waves would otherwise carry into the next wave.

\subsection{Multi-GPU Execution in One Process}
\label{sec:multigpu}

\sysname\ drives all $G$ GPUs of a server from one host thread in one process. Each GPU holds a $1/G$ share of the copies and answers every query of the batch with them, which runs the first round of each query on all GPUs at once. A peer-gather kernel inside the CUDA graph then merges the start points of each query from the other GPUs over NVLink, after which each GPU runs the waves for $1/G$ of the queries. With CUDA events ordering the two graphs of all GPUs without host barriers, the host issues the whole batch in one pass over the GPUs. The wave graph of each GPU starts only after the first-round graphs of all GPUs have recorded their events, which keeps the exchange of start points out of the critical path of the host. Small datasets can instead replicate the whole index on every GPU, which then answers its own share of the queries without exchange.

\sysname\ splits the sketch rows and the neighbor rows by rows over the $G$ GPUs and reads them through peer pointers. The half-precision vectors are split in the same way when they would otherwise crowd out copies. Every kernel receives an array of device pointers to the row ranges of each split part. A row can live on the local HBM, on the HBM of a peer GPU, or in host memory, since the unified address space of CUDA gives all three memories pointers of the same form.

\subsection{Tiered Placement by Access Frequency}
\label{sec:tiers}

\sysname\ places every part of its index by how often a query reads it. The first round reads $C r$ list entries per query. Every wave reads the sketch rows and the neighbor rows of thousands of entries at random positions. These parts stay in HBM, which serves random reads of short rows two orders of magnitude faster than host memory over PCIe. When the copies, sketch rows, and neighbor rows do not fit together, \sysname\ shrinks them instead of moving any of them to the host. The sketch kernels accept any width that is a multiple of 16 and leave the extra lanes of the last group idle. Each copy can also keep only the first entries of every list, which are the nearest to the center of the fragment.

The final rerank reads only $R$ half-precision vectors per query and is the only step that reads host memory. When these vectors do not fit into the HBM that the other parts leave, \sysname\ keeps them in host memory. It allocates them as anonymous memory and fills them from the file in sequential chunks, without a mapping of the file.

\begin{figure}[t]
\centering
\includegraphics[width=\columnwidth]{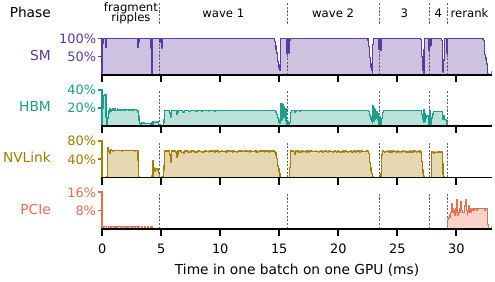}
\caption{GPU utilization in one Deep1B batch by phase.}
\label{fig:monitor}
\end{figure}

With this placement, PCIe carries only the final rerank, and every round before it stays on the HBM and NVLink of the GPUs. To verify this split, we sample the hardware counters of the four H100 GPUs at 20~kHz over 30 batches of 10,000 queries on Deep1B with 512 start points and four waves. Figure~\ref{fig:monitor} plots the raw samples of one batch on one GPU. Through the fragment ripples and all four waves, the SMs stay busy and the GPU reads the sketch rows and neighbor rows of its peers over NVLink at up to 65\% of its bandwidth. PCIe stays at most 1\% busy in these phases. The final rerank alone reads host memory, for 3.5~ms of the 33-ms batch at up to 13\% of the PCIe bandwidth. HBM serves the short random rows at about 18\% of its read bandwidth, as these reads wait on latency rather than bandwidth.

\subsection{Fast Build by Sorting and Sketch Ranking}
\label{sec:fastbuild}

The build of a billion-vector index spends its time on two nearest-neighbor problems: the fragment ripples of hundreds of copies and the neighbor row of every vector. Both grow with the dataset, the first also with the number of copies. \sysname\ solves the first by sorting instead of searching and the second by ranking without full vectors, and it hides the I/O of the build behind the GPU work.

The fragment ripples need no search over the dataset. The pass that hashes the vectors of a copy also computes the projected distance of every vector to the center of its fragment. A radix sort by this distance, followed by a stable radix sort by fragment key, orders the members of every fragment by distance, and each stored fragment keeps its first $l$ members as its ripple. A copy thus costs one streaming pass over the dataset and two sorts per spacing. At one billion vectors, every stored fragment holds at least $p_{\min} = 100$ members, many more than the $l = 16$ entries of a ripple. Smaller datasets hold few vectors per fragment and instead search their ripples with the IVF-Flat index of cuVS~\cite{nvidia2025cuvs}.

At one billion vectors, the neighbor rows come from two passes of the search of \sysname\ over all base vectors: the first round alone, then two waves over the rows of the first pass (\S\ref{sec:waves}). Both passes rank by sketch distance only and skip the final rerank, so that they read no half-precision vectors from host memory. The builder loads the index once for both passes and replaces only the neighbor rows between them. It writes the lists and the sketch in a background thread while the passes run and page-locks the dataset in place instead of copying it into new pinned memory.

\section{Evaluation}
\label{sec:eval}

\subsection{Experimental Setup}

\begin{table}[t]
\centering
\caption{Index configuration of \sysname\ per dataset.}
\label{tab:config}
\vspace{3pt}
\scriptsize
\setlength{\tabcolsep}{1.6pt}
\renewcommand{\arraystretch}{1.15}
\begin{tabular*}{\columnwidth}{@{\extracolsep{\fill}}l*{9}{c}}
\toprule
 & \rotatebox{90}{SIFT} & \rotatebox{90}{GIST} & \rotatebox{90}{DINOv2} & \rotatebox{90}{MARCO} & \rotatebox{90}{D-10M} & \rotatebox{90}{D-100M} & \rotatebox{90}{S-100M} & \rotatebox{90}{D-1B} & \rotatebox{90}{S-1B} \\
\midrule
Copies $C$ & 200 & 400 & 600 & 256 & 320 & 272 & 224 & 256 & 256 \\
List length $l$ & 32 & 64 & 32 & 64 & 64 & 64 & 64 & 16 & 16 \\
Min.\ fragment $p_{\min}$ & 1 & 1 & 2 & 2 & 2 & 10 & 10 & 100 & 100 \\
Sketch width & 64 & 128 & 128 & 128 & 64 & 64 & 64 & 96 & 96 \\
Row length $L_t$ & 64 & 64 & 64 & 64 & 64 & 64 & 64 & 32 & 32 \\
Reverse edges & -- & -- & \checkmark & \checkmark & \checkmark & \checkmark & \checkmark & \checkmark & \checkmark \\
Neighbor rows & exact & exact & exact & exact & exact & search & search & search & search \\
\bottomrule
\end{tabular*}
\end{table}

\begin{table}[t]
\centering
\caption{Baselines and their placement on the four GPUs.}
\label{tab:baselines}
\vspace{3pt}
\scriptsize
\setlength{\tabcolsep}{2.5pt}
\renewcommand{\arraystretch}{1.15}
\begin{tabular*}{\columnwidth}{@{\extracolsep{\fill}}ll*{9}{c}}
\toprule
System & Placement & \rotatebox{90}{SIFT} & \rotatebox{90}{GIST} & \rotatebox{90}{DINOv2} & \rotatebox{90}{MARCO} & \rotatebox{90}{D-10M} & \rotatebox{90}{D-100M} & \rotatebox{90}{S-100M} & \rotatebox{90}{D-1B} & \rotatebox{90}{S-1B} \\
\midrule
\multicolumn{11}{l}{\emph{Graph indexes}} \\
CAGRA~\cite{ootomo2024cagra} & full; 8 shards at 1B & \checkmark & \checkmark & \checkmark & \checkmark & \checkmark & \checkmark & \checkmark & \checkmark & \checkmark \\
GGNN~\cite{groh2023ggnn} & 4 shards; full, $\times$4 & \checkmark &  &  & \checkmark & \checkmark & \checkmark & \checkmark &  &  \\
FlowANN~\cite{zhao2026flowann} & one GPU, $\times$4 &  &  &  &  & \checkmark & \checkmark &  &  &  \\
PathWeaver~\cite{kim2025pathweaver} & 4 shards, end to end &  &  &  &  & \checkmark & \checkmark &  &  &  \\
\midrule
\multicolumn{11}{l}{\emph{Inverted file indexes}} \\
IVF-PQ~\cite{jegou2011pq,nvidia2025cuvs} & full per GPU & \checkmark & \checkmark & \checkmark & \checkmark &  & \checkmark & \checkmark &  &  \\
IVF-RaBitQ~\cite{shi2026ivfrabitq} & full; 8 shards at 1B &  &  &  & \checkmark & \checkmark & \checkmark & \checkmark & \checkmark & \checkmark \\
\bottomrule
\end{tabular*}
\end{table}

\begin{figure*}[t]
\centering
\includegraphics[width=\textwidth]{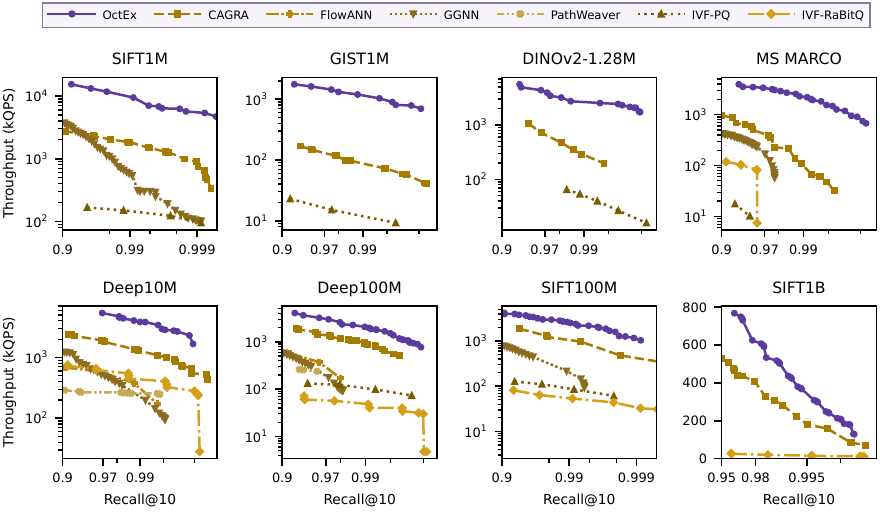}
\caption{Recall and throughput on four H100 GPUs (Pareto frontiers, median QPS, recall axis in nines).}
\label{fig:frontier}
\end{figure*}

\paragraph{Datasets.} We evaluate on nine datasets of image descriptors, image embeddings, and text embeddings, from one million to one billion vectors of 96 to 960 dimensions. SIFT1M (SIFT)~\cite{jegou2011bigann,lowe2004sift} holds one million 128-dimensional SIFT descriptors with 10,000 queries. GIST1M (GIST)~\cite{oliva2001gist,aumuller2020annbench} holds one million 960-dimensional GIST descriptors from ANN-Benchmarks with 1,000 queries. DINOv2-1.28M (DINOv2) holds the normalized 768-dimensional DINOv2~\cite{oquab2024dinov2} embeddings of the 1.28 million ImageNet training images~\cite{russakovsky2015ilsvrc}, with 10,000 validation images as queries. MS MARCO (MARCO)~\cite{nguyen2016msmarco} holds the 8.8 million passages and the 6,980 labeled dev queries of its passage-ranking task as normalized 768-dimensional embeddings of the Sentence-BERT~\cite{reimers2019sbert} model msmarco-distilbert-cos-v5~\cite{sanh2019distilbert}, with ground truth from a single-precision screen of 128 candidates and a double-precision rerank. Deep10M, Deep100M, and Deep1B (D-10M, D-100M, D-1B) hold the first ten million, the first one hundred million, and all one billion 96-dimensional vectors of Deep1B~\cite{babenko2016deep} with its 10,000 queries, with exact ground truth for the two prefixes. SIFT100M and SIFT1B (S-100M, S-1B) hold the first one hundred million and all one billion SIFT descriptors of BIGANN~\cite{jegou2011bigann} with its 10,000 queries and public ground truth.

\paragraph{Configurations of \sysname.} All datasets share one lattice configuration: $b = 4$ blocks of $E_8$ over the top 64 principal coordinates and five spacings from 2.0 to 5.0 projected standard deviations. Table~\ref{tab:config} lists the parts of the index that scale with the dataset. The read depth, the start points, and the waves vary along the frontier of each dataset. The two billion-vector indexes come from the build of Section~\ref{sec:fastbuild} and rerank $R = 32$ entries at the end, while smaller datasets rerank 128 to 256 entries exactly per wave.

\paragraph{Baselines.} We compare \sysname\ with four GPU graph systems and two inverted file indexes, each swept over its own parameters. Table~\ref{tab:baselines} lists the baselines from top to bottom with their placement on the four GPUs. CAGRA~\cite{ootomo2024cagra} in cuVS~\cite{nvidia2025cuvs} 26.10 sweeps graph degrees 32 to 96 (32 at one billion vectors), internal top-$k$ sizes 32 to 1,024, and search widths 1 to 4, in single and half precision. GGNN~\cite{groh2023ggnn} runs with the default parameters of its authors, sharded over the four GPUs and with the full index on one GPU, credited with ideal 4$\times$ scaling. FlowANN~\cite{zhao2026flowann} runs on one GPU with the scripts, settings, and GDRCopy driver of its authors, credited with ideal 4$\times$ scaling, up to one hundred million vectors, as its graph build on Deep1B processes only 6.7k vectors per second. PathWeaver~\cite{kim2025pathweaver} runs its own preprocessing and multi-GPU search with one shard per GPU and is timed end to end, since its own timer starts after the first and most expensive round. IVF-PQ~\cite{jegou2011pq,nvidia2025cuvs} in cuVS sweeps 16 to 256 probes over 2,048 lists at one million and 16,384 lists at one hundred million vectors with a refinement ratio of 4. IVF-RaBitQ~\cite{shi2026ivfrabitq} in cuVS sweeps 4 and 7 bits and refinement ratios 2 to 8 over 4,096 to 65,536 lists, followed by our exact GPU rerank.

\paragraph{Metrics.} We report throughput in QPS over the full query set in batches and recall at 10, which counts the fraction of the true ten nearest neighbors among the ten returned. Throughput is the median over five timed repetitions after a warm-up, for \sysname\ and for every baseline. The median matters for CAGRA at one billion vectors, whose best repetition exceeds its median by up to 1.7$\times$, whereas the best and the median of \sysname\ differ by at most 4.5\%. For each system, we sweep its configurations and report the Pareto frontier of recall and throughput. A speedup at a recall target compares the best throughput of each system at or above that target. A range of speedups over CAGRA spans the targets 0.95, 0.97, 0.98, 0.99, and 0.995 that both systems reach.

\paragraph{Platform.} All measurements outside Section~\ref{sec:eval-a100} run on one server, on all four of its GPUs unless Table~\ref{tab:baselines} notes one. The server holds four NVIDIA H100 GPUs with 93~GiB of HBM each, connected all to all by NVLink, Intel Xeon Platinum 8468 processors with 192 hardware threads, and 960~GB of DRAM, with CUDA 12.6 and cuVS 26.10. Its datasets and indexes reside on a network-attached block device that reads and writes about 0.8~GB/s per stream and reads 6.7~GB/s with eight streams in parallel. Section~\ref{sec:eval-a100} uses a second server with one whole NVIDIA A100 80GB PCIe GPU, two AMD EPYC 7763 processors with 256 threads, and 512~GB of DRAM.

\begin{figure*}[t]
\centering
\includegraphics[width=\textwidth]{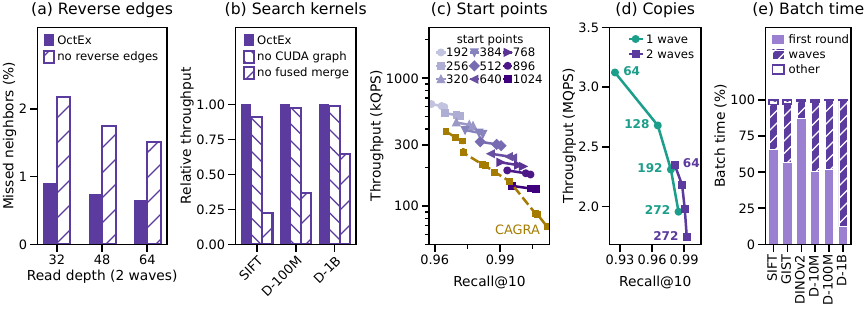}
\caption{Reverse edges, system optimizations, start points, copies, and the batch time by phase.}
\label{fig:ablation}
\end{figure*}

\subsection{Recall--Throughput Trade-off}

We compare \sysname\ with the baselines over the full recall range on all nine datasets, to test whether a fixed number of parallel steps turns HBM into throughput at high recall. Each Pareto frontier credits every system with its best configuration of Tables~\ref{tab:baselines} and~\ref{tab:config} at each recall.

Figures~\ref{fig:frontier} and~\ref{fig:opener}(a) show that \sysname\ leads every baseline at every recall that both reach, on all nine datasets. Up to one hundred million vectors, it outperforms CAGRA by 2.0$\times$ to 46.4$\times$ from recall 0.95 to 0.995. Over the other GPU graph systems and the inverted file indexes, the gain reaches 5.4$\times$ to 27.9$\times$ and 7.0$\times$ to 343$\times$. The gain grows with the recall target, for instance from 5.9$\times$ at recall 0.95 to 46.4$\times$ at 0.995 on MS MARCO, where CAGRA quadruples its candidate list to rise from recall 0.987 to 0.996. It narrows with the size of the dataset, to 2.0$\times$ to 2.5$\times$ on Deep100M, as a larger dataset needs more copies and waves to reach the same neighbors. Figure~\ref{fig:opener}(a) shows the same comparison at one billion vectors, where the index of \sysname\ spans the HBM of all four GPUs and the host memory. There, \sysname\ still outperforms CAGRA by 1.48$\times$ to 2.05$\times$ and IVF-RaBitQ by 20.2$\times$ to 26.7$\times$. On SIFT1B in Figure~\ref{fig:frontier}, the gain over CAGRA is 1.46$\times$ to 1.85$\times$ and reaches 2.38$\times$ at recall 0.998, against 20.8$\times$ to 38.9$\times$ over IVF-RaBitQ. The gain of \sysname\ is thus largest at high recall, where graph search needs its longest chains of rounds.

\subsection{Ablation and Sensitivity}
\label{sec:eval-ablation}

We ablate two design choices of \sysname, the reverse edges of the neighbor rows and the system optimizations of the search, and vary the two parameters that set its reach, the start points and the copies. Each panel of Figure~\ref{fig:ablation} varies one of them on the largest dataset at which it matters.

\begin{figure*}[t]
\centering
\includegraphics[width=\textwidth]{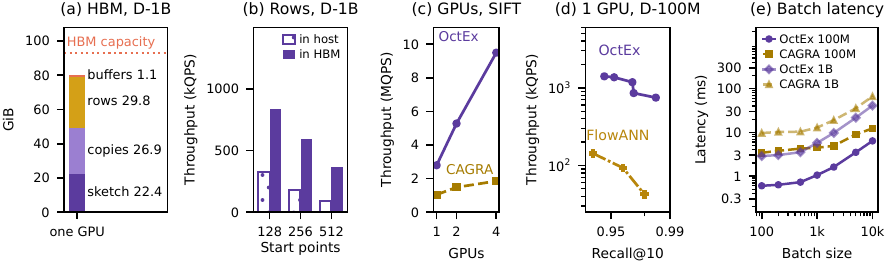}
\caption{Placement at one billion vectors, the neighbor-row tier, GPU scaling, one GPU against FlowANN, and batch latency.}
\label{fig:placement}
\end{figure*}

Both design choices pay off: the reverse edges halve the missed neighbors at a small cost, and the fused merge kernels and the CUDA graphs raise the throughput on every dataset. Figure~\ref{fig:ablation}(a) ablates the reverse edges of the neighbor rows on Deep100M with two waves. A row without reverse edges lists only the nearest vectors of its owner, whereas a row of \sysname\ of the same length also lists vectors that count the owner among their own nearest. The reverse edges cut the missed neighbors from 2.2\% to 0.9\% at read depth 32 and by more than half at every read depth, while the throughput drops by only 2.6\% to 3.7\%. Figure~\ref{fig:ablation}(b) ablates the two search kernel optimizations one at a time and reports the throughput relative to the full search at the same recall. Without the fused merge kernels, which select the start points in shared memory, the merge must sort all candidates in global memory. The throughput then falls to 23\% on SIFT1M, 37\% on Deep100M, and 65\% on Deep1B. Without CUDA graphs, the host launches every kernel and synchronizes between the first round and the waves, which costs 9\% on SIFT1M, whose batches take only 1.4~ms, and at most 2.3\% on the larger datasets.

The start points, the copies, and the waves set the reach of \sysname, and at one billion vectors the waves take most of its time. Figure~\ref{fig:ablation}(c) varies the start points on Deep1B, each line from three to five waves. Every further increase extends the frontier toward higher recall. With the waves fixed at five, tuning only the start points loses at most 1.9\% of throughput against the best configuration at every recall target from 0.97 to 0.995, which makes them the main tuning parameter of a loaded index. Figure~\ref{fig:ablation}(d) varies the copies on Deep100M and shows that copies and waves can replace each other: 64 copies with two waves reach both higher recall and higher throughput than one wave over 192 copies. Figure~\ref{fig:ablation}(e) splits the batch time by phase. The waves, which include the final rerank, take 12\% to 48\% of it up to one hundred million vectors but 88\% on Deep1B. As a result, the cost of a wave decides the throughput at one billion vectors.

\subsection{Robustness to Resource Limits}
\label{sec:eval-deploy}

We test whether the gains of \sysname\ hold when its index exceeds the HBM, when fewer GPUs are available, and when batches shrink. The first two panels of Figure~\ref{fig:placement} cover the memory tiers of Deep1B, and the last three cover GPU scaling, a single GPU, and the batch size.

At one billion vectors, \sysname\ stays fast as long as every row that a wave reads stays in HBM. Figure~\ref{fig:placement}(a) shows the memory of one GPU at one billion vectors. Each GPU holds a quarter of the copies, sketch rows, and neighbor rows together with its search buffers in 80.2 of its 93~GiB, while the 179~GiB of half-precision vectors stay in host memory for the final rerank. Lists of 32 entries do not fit 256 copies next to these rows; at 128 copies, keeping only their first 16 entries cuts the list memory from 93.8 to 53.8~GiB at a recall loss of 0.0012. Figure~\ref{fig:placement}(b) moves the neighbor rows to host memory as well. The throughput then falls by 2.6$\times$ to 3.9$\times$, as GPU reads of random rows from host memory reach only 40 million rows per second per GPU against 5.4 billion from HBM.

\sysname\ keeps its gains with fewer GPUs and with smaller batches. Figure~\ref{fig:placement}(c) scales SIFT1M from one to four GPUs, each with the full index and a share of the queries. The throughput of \sysname\ grows by 3.4$\times$ to 9.50 million QPS at recall 0.991, against 1.8$\times$ for CAGRA, since every GPU of \sysname\ runs the CUDA graphs of a batch without host synchronization. Figure~\ref{fig:placement}(d) compares \sysname\ with FlowANN on one H100 with Deep100M, the single-GPU setting of FlowANN, where \sysname\ outperforms it by 12.8$\times$ to 27.9$\times$ from recall 0.95 to 0.98. Figure~\ref{fig:placement}(e) shrinks the batch from 10,000 to 100 queries near recall 0.99. On Deep100M, the batch time of \sysname\ falls from 6.4 to 0.60~ms, as a batch runs a fixed number of parallel steps, whereas that of CAGRA falls only from 12.3 to 3.4~ms. On Deep1B, the gain of \sysname\ consequently grows from 1.6$\times$ at 10,000 queries per batch to 3.4$\times$ at 100 queries.

\subsection{Build Time and Memory Footprint}
\label{sec:eval-build}

\begin{figure}[t]
\centering
\includegraphics[width=\columnwidth]{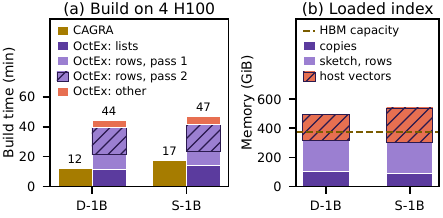}
\caption{Build time and memory at one billion vectors.}
\label{fig:build}
\end{figure}

\sysname\ buys its throughput with precomputation: hundreds of copies whose lists are computed at build time and held in HBM. We measure this cost at one billion vectors, in the time to build the index and in the memory that the loaded index occupies. Both costs decide which datasets suit \sysname, namely those that are indexed once and queried often.

Figure~\ref{fig:build}(a) shows that a billion-vector index of \sysname\ takes less than an hour to build on four H100 GPUs, each of which builds a quarter of the copies without communication. \sysname\ builds Deep1B in 44 minutes and SIFT1B in 47 minutes, against 12 and 17 minutes for the eight shards of CAGRA. The fragment ripples of the 256 copies take 12 and 14 minutes, one streaming pass over the dataset per copy. The two passes of the neighbor rows together take another 28 minutes, as each of them searches all one billion base vectors. Reading the dataset and computing the sketch take the remaining minutes, while the index files are written during the passes.

Figure~\ref{fig:build}(b) shows that memory rather than build time bounds the index of \sysname\ at one billion vectors. On Deep1B, the sketch rows and the neighbor rows hold 66\% of the 316~GiB in HBM, against 34\% for the 256 copies with lists of 16 entries. Deep1B and SIFT1B add 179 and 238~GiB of half-precision vectors in host memory, which only the final rerank reads. The HBM of the server thus bounds the index, which is why both datasets keep only 16 entries per list (Figure~\ref{fig:placement}(a)).

\subsection{Portability to Other GPU Platforms}
\label{sec:eval-a100}

\begin{figure}[t]
\centering
\includegraphics[width=\columnwidth]{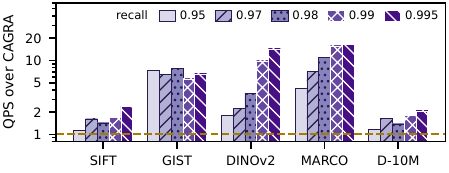}
\caption{Speedup over CAGRA on one A100 by recall target.}
\label{fig:a100}
\end{figure}

We repeat the comparison with CAGRA on the whole A100 of the previous generation. \sysname\ keeps its configurations with as many copies as one GPU holds, while CAGRA sweeps degrees 32 to 96 with internal top-$k$ sizes up to 1,024. With 80~GB of HBM on one GPU, \sysname\ loads all 200 copies of SIFT1M and all 600 of DINOv2-1.28M, but only 112 of the 400 copies of GIST1M, 69 of the 256 of MS MARCO, and 87 of the 320 of Deep10M. Both systems run the same CUDA 12.6 and cuVS 26.10 as on the H100 server.

Figure~\ref{fig:a100} shows that \sysname\ outperforms CAGRA on all five datasets at every recall target from 0.95 to 0.995, by 1.14$\times$ to 2.40$\times$ on SIFT1M, 5.62$\times$ to 7.89$\times$ on GIST1M, 1.81$\times$ to 14.4$\times$ on DINOv2-1.28M, 4.20$\times$ to 16.3$\times$ on MS MARCO, and 1.19$\times$ to 2.08$\times$ on Deep10M. As on the H100, the gains are largest on high-dimensional vectors and at the highest recall targets. Loaded with only 8 to 32 of its copies, which take 1.7$\times$ to 3.1$\times$ the memory of CAGRA, \sysname\ still outperforms CAGRA on MS MARCO by 2.1$\times$ at recall 0.95, 4.7$\times$ at recall 0.98, and 6.3$\times$ at recall 0.99.

\section{Related Work}
\label{sec:related}

\paragraph{Graph and inverted file indexes.} Proximity-graph indexes such as HNSW~\cite{malkov2020hnsw}, NSG~\cite{fu2019nsg}, and DiskANN~\cite{subramanya2019diskann} expand the best unvisited vertices of a query round by round. CAGRA~\cite{ootomo2024cagra}, GGNN~\cite{groh2023ggnn}, SONG~\cite{zhao2020song}, and BANG~\cite{venkatasubba2025bang} run this loop on GPUs. FlowANN~\cite{zhao2026flowann} tiers a CAGRA graph across GPU and host memory, whereas PathWeaver~\cite{kim2025pathweaver} pipelines a sharded CAGRA search across GPUs. PipeANN~\cite{pipeann2025} aligns the same loop with SSD reads. Their graphs are neighbor tables refined in the spirit of NN-Descent~\cite{dong2011nndescent}, with the reverse links that HNSW~\cite{malkov2020hnsw} and NSG~\cite{fu2019nsg} add for connectivity. Inverted file indexes in Faiss~\cite{johnson2019faiss,douze2024faisslib} and cuVS~\cite{nvidia2025cuvs} instead scan the codes of the probed lists, compressed by product quantization~\cite{jegou2011pq,ge2013opq}, inverted multi-indexes~\cite{babenko2012imi}, or RaBitQ~\cite{gao2024rabitq,shi2026ivfrabitq}. Milvus~\cite{wang2021milvus}, Manu~\cite{guo2022manu}, Starling~\cite{wang2024starling}, FusionANNS~\cite{tian2025fusionanns}, and Helmsman~\cite{helmsman} distribute or tier such indexes across machines, memory, and storage. SPANN~\cite{chen2021spann} and SOAR~\cite{sun2023soar} spend memory on recall by assigning a vector to several lists, but every query still scans the lists that it probes. \sysname\ reads the same kind of neighbor table as a graph walk and spends memory as SPANN and SOAR do, yet it expands all start points of a query at once in a fixed number of waves and reads precomputed lists instead of scanning partitions.

\paragraph{Random projections.} Locality-sensitive hashing~\cite{indyk1998ann,datar2004lsh,andoni2008nearoptimal} and its multi-probe variants~\cite{lv2007multiprobe} hash vectors into the cells of random projections, where Jégou et al.~\cite{jegou2008qalsh} use the $E_8$ lattice as the quantizer. C2LSH~\cite{gan2012c2lsh} and QALSH~\cite{huang2015qalsh} count collisions over one-dimensional projections, as SuCo~\cite{suco2025} does over subspaces. GPU implementations~\cite{pan2011gpulsh,wang2018flash} accelerate the bucket lookups of such indexes. All of them store which vectors fall into each bucket and return the members of the buckets of a query, whose number varies with the population of each bucket. \sysname\ keeps the random projections and the $E_8$ quantizer of these indexes~\cite{jegou2008qalsh} but reads a fixed number of entries per copy, which gives every query the same work on the GPU.

\paragraph{Approximate Voronoi diagrams.} The approximate Voronoi diagrams of Har-Peled~\cite{harpeled2001avd} and Arya et al.~\cite{arya2009spacetime} store an answer per cell of a partition, so that a query reads the answer of its own cell instead of searching. Their number of cells, however, grows exponentially with the dimension. \sysname\ stores such answers only in low-dimensional projections: for every populated fragment, the list of its nearest vectors, which reaches beyond the fragment by the radius of Section~\ref{sec:ripples}. Its waves then reach the neighbors that no copy holds, which no stored answer per cell provides.

\paragraph{Scheduling irregular search on GPUs.} Irregular graph workloads face the same imbalance as a query whose lists differ in length. Hong et al.~\cite{hong2011warp} assign whole adjacency lists to warps, whereas Merrill et al.~\cite{merrill2012gputraversal}, Davidson et al.~\cite{davidson2014sssp}, and Gunrock~\cite{wang2015gunrock} spread the lists of a frontier evenly over the threads by prefix sums and searches, in the manner of merge path~\cite{green2012mergepath}. The gather kernel of \sysname\ follows the second approach for the lists of a query. Selecting the best candidates is a second recurring step, which Faiss performs by warp-level $k$-selection~\cite{johnson2019faiss} and Dr.~Top-k by delegate-centric selection~\cite{gaihre2021drtopk}. The fused merge kernels of \sysname\ instead build on GPU hash tables~\cite{ashkiani2018hashtable}, radix sorting~\cite{satish2009gpusort,adinets2022onesweep}, and bitonic networks~\cite{batcher1968sorting}, which also remove duplicate candidates. Persistent-thread designs~\cite{gupta2012persistentthreads} keep one kernel resident and schedule the work inside it. Nimble~\cite{kwon2020nimble} schedules the kernels of deep learning models ahead of time to cut their launch overhead. With fixed kernel shapes, \sysname\ goes further and replays the whole search of a batch as CUDA graphs~\cite{nvidia2025cudagraphs}, whereas persistent kernels and per-query launches decide their work at run time.

\section{Conclusion}
\label{sec:conclusion}

In this paper, we present \sysname, a GPU index that reaches high recall in a fixed number of parallel steps.
Our design is inspired by the observation that a query's nearest neighbors are often found in the same fragments of a lattice, which we exploit by storing the nearest vectors of every fragment as ripples.
\sysname\ replaces the search rounds with stored ripples: a query reads the precomputed lists of its lattice fragments in hundreds of random projections at once and then the neighbor rows of its best hits in a few waves. 
We open-source \sysname\ as a CUDA module that builds the index of one billion vectors in 44 to 47 minutes, runs each batch as CUDA graphs over all GPUs, and moves only the vectors of the final rerank to host memory;
moreover, we provide Python and C interfaces and drop-in replacements for Faiss and cuVS to facilitate its integration into existing systems.
We evaluate \sysname\ on nine datasets of up to one billion vectors, where it outperforms the state-of-the-art GPU graph systems and inverted file indexes at every recall from 0.95 to 0.995, by up to 46.4$\times$ over CAGRA and by 5.4$\times$ to 343$\times$ over the other baselines.

{\footnotesize \bibliographystyle{acm}
\bibliography{refs}}

\end{document}